\documentstyle[psfig]{mn}
\def\ale{\mathrel{\hbox{\rlap{\hbox{\lower4pt\hbox{$\sim$}}}\hbox{$<$}}}}
\def\age{\mathrel{\hbox{\rlap{\hbox{\lower4pt\hbox{$\sim$}}}\hbox{$>$}}}}
 
\begin{document}

\title[The Spatial Distribution of NS--NS mergers] {The Spatial
Distribution of Coalescing Neutron Star Binaries: Implications for
Gamma-Ray Bursts}

\author[Bloom, Sigurdsson, \& Pols]{Joshua
S.~Bloom$^{1,2}$ and Steinn Sigurdsson$^1$, and Onno R.~Pols$^{1,3}$
 \\ $^1$
Institute of Astronomy, Madingley Road, Cambridge, CB3 0HA, England \\
$^2$ California Institute of Technology, MS 105-24, Pasadena, CA 91106
USA \\ 
$^3$ Instituto de Astrof\'{\i}sica de Canarias, c/ Via L\'actea s/n, 
E-38200 La Laguna, Tenerife, Spain \\
email: {\tt jsb@astro.caltech.edu}}
 
\date{}
 
\maketitle
 
\begin{abstract}

We find the distribution of coalescence times, birthrates, spatial
velocities, and subsequent radial offsets of coalescing neutron stars
(NSs) in various galactic potentials accounting for large asymmetric
kicks introduced during a supernovae.  The birthrates of bound NS--NS
binaries are quite sensitive to the magnitude of the kick velocities
but are, nevertheless, similar ($\sim 10$ per Galaxy per Myr) to
previous population synthesis studies.  The distribution of merger
times since zero-age main sequence is, however, relatively insensitive
to the choice of kick velocities.  With a median merger time of $\sim
10^8$ yr, we find that compact binaries should closely trace the star
formation rate in the Universe.

In a range of plausible galactic potentials (with $M_{\rm galaxy} \age
3\times 10^{10} M_\odot$) the median radial offset of a NS--NS mergers
is less than 10 kpc. At a redshift of $z=1$ (with $H_0 = 65$ km
s$^{-1}$ Mpc$^{-1}$ and $\Omega = 0.2$), this means that half the
coalescences should occur within $\sim 1.3$ arcsec from the host
galaxy.  In all but the most shallow potentials, ninety percent of
NS--NS binaries merge within 30 kpc of the host.  We find that although
the spatial distribution of coalescing neutron star binaries is
consistent with the close spatial association of known optical
afterglows of gamma-ray bursts (GRBs) with faint galaxies, a
non-negligible fraction ($\sim 15$ percent) of GRBs should occur well
outside ($\age 30$ kpc) dwarf galaxy hosts.  Extinction due to dust in
the host, projection of offsets, and a range in interstellar medium
densities confound the true distribution of NS--NS mergers around
galaxies with an observable set of optical transients/galaxy offsets.
\end{abstract}


\section{Introduction}

The discovery of an X-ray afterglow by BeppoSAX (Costa et al.~1997;
Piro, Scarsi, \& Butler~1995) and subsequently an optical transient
associated with gamma-ray burst (GRB) 970228 (van Paradijs et
al.~1997) led to the confirmation of the cosmological nature of GRBs
(Metzger et al.~1997).  The broadband optical afterglow has been
modeled relatively successfully (M\'esz\'aros \& Rees 1993; Wijers,
Rees, \& M\'esz\'aros~1997; Waxman 1997; Waxman, Kulkarni \& Frail 1998)
as consistent with an expanding relativistic fireball (Rees \&
M\'esz\'aros 1994; Paczy\'nski \& Rhoads 1993; Katz 1994; M\'esz\'aros
\& Rees 1997; Vietri 1997; Sari, Piran, \& Narayan 1998; Rees \&
M\'esz\'aros 1998).

Still, very little is known about the nature of the progenitors of
GRBs, and, for that matter, their hosts.  Broad-band fluence measures
and the known redshifts of some bursts implies a minimum (isotropic)
energy budget for GRBs of $\sim 10^{52-53}$ ergs (Metzger et al.~1997;
Kulkarni et al.~1998; Djgorovski et al.~1998).  The log $N$-log $P$
brightness distribution, the observed rate, $N$, of bursts above some
flux, $P$, versus flux, indicates a paucity of dim events from that
expected in a homogeneous, Euclidean space.  With assumptions of a
cosmology, source evolution and degree of anisotropy of emission, the
log $N$-log $P$ has been modeled to find a global bursting rate.
Assuming the bursts are non-evolving standard candles Fenimore \&
Bloom 1995 found $\sim 1$ bursts events per galaxy per Myr (GEM) to be
consistent with the observed log $N$-log $P$.  More recently, Wijers
et al.~1998 (see also, Totani et al.~1998; Lipunov, Postnov, \&
Prokhorov 1997) found the same data consistent with GRBs as standard
candles assuming the bursting rate traces the star-formation rate
(SFR) in the Universe; such a distribution implies a local burst rate
of $\sim 0.001$ GEM and a standard peak luminosity of $L_0 = 8.3
\times 10^{51}$ erg s$^{-1}$ (Wijers et al.~1998).

Given the energetics, burst rate and implied fluences, the
coalescence, or merger, of two bound neutron stars (NSs) is the
leading mechanism whereby gamma-ray bursts are thought to arise
(Paczy\'nski 1986; Goodman~1986; Eichler et al.~1989; Narayan,
Paczy\'nski, \& Piran 1992).  One quantifiable prediction of the
NS--NS merger hypothesis is the spatial distribution of GRBs (and GRB
afterglow) with respect to their host galaxies. Conventional wisdom,
using the relatively long--lived Hulse-Taylor binary pulsar as a
model, is that such mergers can occur quite far ($\age 100$ kpc)
outside of a host galaxy.  Observed pulsar (PSR) binaries with a
NS companion provide the only direct constraints on such a
populations, but the observations are biased both towards long lived
systems, and systems that are close to the Galactic plane.

The merger rate of NS--NS binaries has been discussed both in the
context of gravitational wave-detection and GRBs (eg.~Phinney 1991;
Narayan et al.~1991; Tutukov \& Yungelson 1994; Lipunov et al.~1995).
Recently Fryer, Burrows, and Benz (1997), Lipunov, Postnov, \&
Prokhorov (1997), Portegies Zwart \& Spreeuw (1996) studied the effect
of asymmetric kicks on birthrates of NS--NS binaries, but did not
quantify the spatial distribution of such binaries around their host
galaxies.  Tutukov \& Yungelson (1994) discussed the spatial
distribution of NS--NS mergers but neglected asymmetric kicks and the
effect of a galactic potential in their simulations.  Only Zwart \&
Yungelson (1998) have discussed the maximum travel distance of merging
neutron stars including asymmetric supernovae kicks.

It is certainly of interest to find the rate of NS--NS coalescences
{\it ab initio} from population synthesis of a stellar
population. This provides an estimate of beaming of GRBs, assuming
they are due to NS--NS mergers, and hence an estimate of probable
frequency of gravitational wave sources, providing a complementary
rate estimate to those of Phinney 1991 and Narayan et al.~1991, which
are based on long lived NS-PSR pairs only and are very
conservative. It also provides an estimate of how the NS--NS mergers
trace the cosmological star formation rate (SFR) of the Universe, if
mean formation rates and binarity of high mass stars are independent
of star formation environments such as metallicity.

Here we concentrate on estimating the spatial distribution of
coalescing NS--NS binaries around galaxies.  To do so, both the system
velocity and the interval between formation of the neutron star binary
and the merger through gravitational radiation is found by simulation
of binary systems in which two supernovae occur.  We explore the
effects of different asymmetric kick amplitudes, and the resultant
birthrates and spatial distribution of coalescing NS--NS binaries born
in different galactic potentials.

In section 2 we briefly outline the prescription for our Monte Carlo
code to simulate bound binary pairs from an initial population of
binaries by including the effect of asymmetric supernovae kicks.  In
section 3 we outline the integration method of NS--NS pairs in various
galactic potentials.  Section 4 highlights the birthrates and spatial
distributions inferred from the simulations.  Section 5 concludes by
discussion the implications and predictions for gamma-ray burst
studies.

\section{Neutron Star Binary Population Synthesis}

We used a modified version of the code created for binary evolution by
Pols (Pols \& Marinus 1994) taking into account the evolution of
eccentricity through tidal interaction and mass transfer before the
first and second supernova, and allowing for an asymmetric kick to
both the NS during supernova. The reader is referred to Pols \&
Marinus (1994) for a more detailed discussion account of the binary
evolution code.

\subsection{Initial Conditions and Binary Evolution}

In general, the evolution of a binary is determined by the initial
masses of the two stars ($m_1$, $m_2$), the initial semi-major axis
($a_o$) and the initial eccentricity ($e_o$) of the binary at zero-age
main sequence (ZAMS).  We construct Monte Carlo ensembles of high-mass
protobinary systems (with primary masses between $4 M_\odot$ and $100
M_\odot$) by drawing from an initial distribution of each of the four
parameters as prescribed and motivated in Portegies Zwart \& Verbunt
(1996).  We treat mass transfer and common-envelope (CE) phases of
evolution as in Pols \& Marinus (1994). CE evolution is treated as a
spiral-in process; we use a value of $\alpha=1$ for the efficiency
parameter of conversion of orbital energy into envelope potential
energy, see eq.~[17] of Pols \& Marinus (1994).  We treat
circularization of an initially eccentric orbit as in Portegies Zwart
\& Verbunt (1996).

During detached phases of evolution we assume that mass accreted by
the companion is negligible so that $a M_{\rm tot}$ = constant.  Mass
lost by the binary system in each successive time step results in a
change in eccentricity according to the sudden mass loss equations
(see, for example, eqns.~[A.21] and [A.24] of Wettig \& Brown
1996). We ignore the effect of gravitational radiation and magnetic
braking in the early stages of binary evolution.

The simple approximation of the 4-parameter distribution function,
albeit rather {\it ad hoc}, appears to adequately reproduce the
observed population of lower mass stars in clusters (eg.~Pols \&
Marinus 1994).  The effect on the distribution of NS--NS binaries
after the second supernova by variation of the 4-parameter space is
certainty of interest, but we have used the canonical values.  A fair
level of robustness is noted in that varying the limits of the initial
distributions of $a_o$ and $e_o$ does not the change the implied
birthrates of bound NS binaries nearly as much as plausible variation
in asymmetric kick distribution.  This effect was noted in Portegies
Zwart \& Spreeuw 1995 and Portegies Zwart \& Verbunt 1995.

\subsection{Asymmetric Supernovae Kicks}

Several authors (eg.~Paczy\'nski 1990; Narayan \& Ostriker 1990; Lyne
\& Lorimer 1994; Cordes \& Chernoff 1997) have sought to constrain the
distribution of an asymmetric kick velocities from observations of
isolated pulsars which are the presumed by-products of type II
supernovae.  Even careful modeling of the selection effects in
observing such pulsars has yielded derived mean velocities that differ
by nearly an order of magnitude.  It is important here to use a good
estimate for the actual physical impulse (the ``kick velocity'') the
neutron stars receive on formation. The observed distribution of
pulsar velocities does not reflect the kick distribution directly as
it includes the Blaauw kick (Blaauw 1961) from those pulsars formed in
binaries, and selection effects on observing both the high and low
speed tail of the pulsar population (eg.~Hartman 1997).  Hansen \&
Phinney (1997) found that the observed distribution is adequately fit
by a Maxwellian velocity distribution with $\sigma_{\rm kick} = 190$
km s$^{-1}$ (which corresponds to a 3-D mean velocity of ~300 km
s$^{-1}$).  Since it is not clear that pulsar observations require a
more complicated kick-velocity distribution, we chose to adopt a
Maxwellian but vary the value of $\sigma_{\rm kick}$.

When a member of the binary undergoes a supernovae we assume the
resulting NS receives a velocity kick, $v_k$, drawn from this
distribution.  Although the direction of this kick may be coupled to
orientation of the binary plane, we choose a kick with a random
spatial direction, since there is no known correlation between the
kick direction or magnitude and the binary parameters.

With an angle $\alpha$ between the velocity kick and the relative
velocity vector, $v$, of the two stars. Then, following earlier
formulae (e.g.~Portegies Zwart \& Verbunt 1996; Wettig \& Brown 1996),
the new-semi major axis of the binary is,
\begin{equation}
a' = \left( \frac{2}{r} - \frac{v^2 + v_k^2 + 2 v v_k \cos \alpha}
	{{\rm G}({\rm M}_{\rm NS} + {\rm M}_2)}\right)^{-1}~.
\label{eq:a}
\end{equation}
where $r$ is the instantaneous distance between the two stars before
SN, $M_2$ is the mass of the companion (which may already be a NS),
and $M_{\rm NS}=1.4 M_{\odot}$ is the mass of the resulting neutron
star.  We neglect the effects of supernova-shell accretion on the mass
of the companion star.  If $a'$ is positive, the new eccentricity is
\begin{equation}
e' = \left[1 - \frac{| \vec r \times \vec v_r|^2}{a'{\rm G}({\rm
M}_{\rm NS} + {\rm M}_2)}\right]^{1/2}~,
\label{eq:e}
\end{equation}
where the resultant relative velocity is $\vec v_r = \vec v + \vec
v_k$. Assuming the kick directions between successive SN are
independent, the resulting kick to the bound system (whose magnitude
is given by equation [2.10] of Brant \& Podsiadlowski 1995) is added
in quadrature to the initial system velocity to give the system
velocity ($v_{\rm sys}$).
 
To produce 1082 bound NS--NS binaries with a Hansen \& Phinney kick
velocity distribution and initial conditions described above, we
follow the evolution of 9.7 million main sequence binaries which
produce a total of $\sim$ 1 million neutron stars through supernovae.
Assuming a supernova rate of 1 per 40 years (Tammann et al.~1994) and
40\% binary fraction (as in Portegies Zwart \& Spreeuw 1996), we find
an implied birthrate of NS--NS binaries by computing the number of
binaries with SN type II per year and multiply by the ratio of bound
NS--NS systems to SN type II as found in the simulations.  We neglect
the (presumed small) contribution of other formation channels (eg.~
three-body interactions) to the overall birthrate of NS--NS binaries.
The implied birthrate of NS--NS binaries from various kick-velocity
magnitudes are given in table 2.


\section{Evolution of binaries systems in a galactic potential}

The large-scale dynamics of stellar objects are dominated by the halo
gravitational potential while the initial distribution of stellar
objects is characterized by a disk scale length.  We take the disk
scale and halo scale to vary independently in our galactic models.  We
assume the NS--NS binaries are born in an exponential stellar disk,
with birthplace drawn from randomly from mass distribution of the
disk. The initial velocity is the local circular velocity
(characterized but the halo) plus $v_{\rm sys}$ added with random
orientation.

We then integrate the motion of the binary in the galactic potential
assuming a Hernquist (1990) halo; we ignore the contribution of the
disk to the potential.  We assume scale lengths for the disk and halo,
the disk scale ($r_{\rm disk}$) determines the disk distribution, the
halo scale length ($r_{\rm break}$) and circular velocity ($v_{\rm
circ}$) determine the halo mass (see table 1).  The movement of the
NS--NS binaries on long time-scales is sensitive primarily to the
depth of the galactic potential (here assumed to be halo dominated)
and how quickly it falls off at large radii.  Assuming isothermal
halos instead of Hernquist profiles would decrease the fraction of
NS--NS pairs that move to large galactocentric radii, but the
differences in distribution are dominated by the true depth of the
halo potentials in which the stars form rather than their density
profiles at large radii.

\begin{figure}
\centerline{\psfig{file=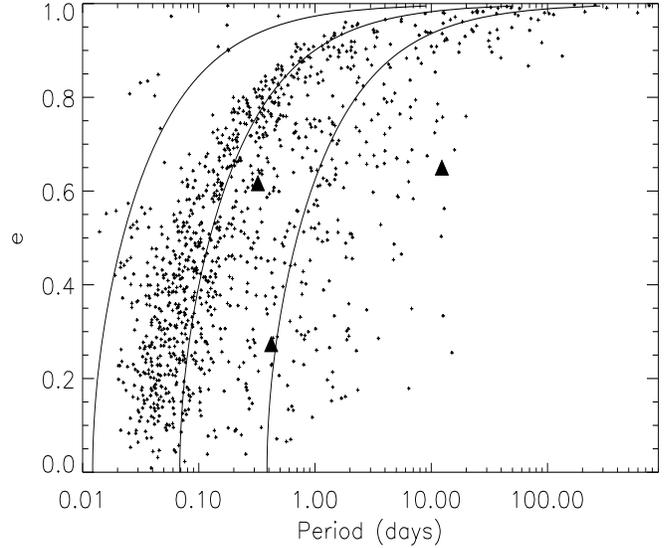,height=3.2in,width=3.9in}}
\caption[]{The distribution of orbital parameters (period and
eccentricity) after the second supernovae for bound NS--NS pairs. From
left to right, are lines of constant merger time after second SN
($10^6$, $10^8$, $10^{10}$ yrs). The parameters of the observed NS
pairs 1913+16 (Taylor \& Weisberg 1989), 1534+12 (Wolszczan 1991), and
2303+46 (Taylor \& Dewey 1988) are marked with triangles.  With an
observational bias towards long-lived systems, clearly the observed
PSR-NS systems are not indicative of the true NS binary distribution
(see section 4.1).}
\label{fig:ae}
\end{figure} 
We use a symplectic leapfrog integrator to advance the binary in the
galactic potential, and a simple iteration scheme to evolve the
semi-major axis and eccentricity of the binary as gravitational
radiation drives $a$ and $e$ to zero, assuming the orbit averaged
quadrupole dominated approximation (Peters 1964).  The integration is
continued until either $1.5 \times 10^{10}$ years have passed (no merger in
Hubble time) or the characteristic time to merger is short compared to
the dynamical time scale of the binary in the halo (ie.~the binary
won't move any further before it merges).  We then record the 3-D
position of the binary relative to the presumptive parent galaxy and
the time since formation.

\section{Results}

\subsection{Orbital parameter distribution after the second supernova}

Figure \ref{fig:ae} shows the distribution of orbital parameters
(semi-major axis and period) after the second supernova for bound
NS--NS pairs for the Hansen \& Phinney (1997) kick distribution
($\sigma_{\rm kick} = 190$ km s$^{-1}$). As found previously
(eg.~Portegies Zwart \& Spreeuw 1996), bound systems tend to follow
lines of constant merger time.  The density of systems in figure 1 can
be taken as the probability density of finding a NS--NS binaries
directly after the second supernova.  In time, the shorter-lived
systems (higher $e$ and shorter period) merge due to gravitational
radiation.  Thus, at any given time after a burst of star-formation
there is an observational bias towards finding longer-lived systems.
In addition, there is a large observational bias against finding short
period binaries (Johnston \& Kulkarni 1991).  That the observed PSR-NS
systems lie in the region of parameter space with low initial
probability is explained by these effects.  The time-dependent
probability evolution has been discussed and quantified in detail by
Portegies Zwart \& Yungelson (1998).
\begin{figure}
\centerline{\psfig{file=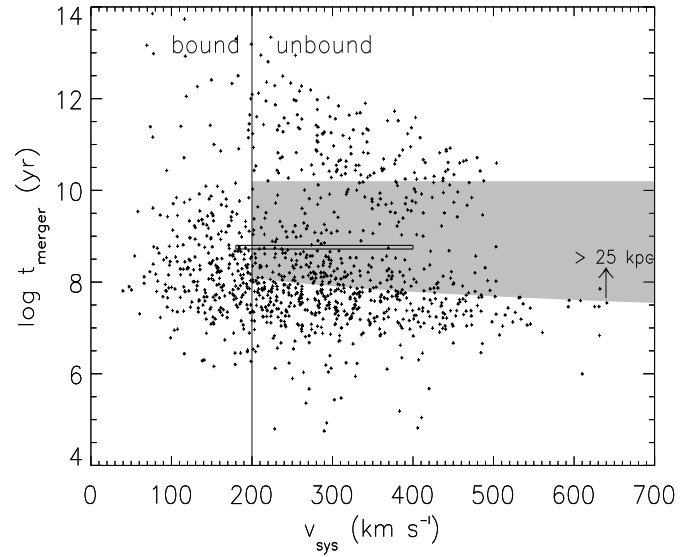,height=3.2in,width=3.9in}}
\caption[]{The distribution of merger times after second supernovae as
a function of system velocity.  Left of the vertical line, all pairs
created are gravitationally bound to a undermassive host ($3 \times
10^{10} M_\odot$) at the disk scale radius.  Of the pairs that are
unbound, only the pairs in the shaded region could travel more than
$\sim 25$ kpc (linearly) from their birthplace and merge within a
Hubble time ($\ale 1.5\times 10^{10}$ yrs). Since the spatial velocity
of observed NS binaries includes both the initial circular velocity of
the system and the system velocity due to kicks from each supernova,
the true system velocities are highly uncertain.  For comparison,
though, we demark the range of accepted kick velocities of PSR 1913+16
with a long rectangle (the merger time is much better constrained than
that depicted); this illustrates a general agreement of the system
velocity of PSR 1913+16 and the modeled distribution of bound NS
binaries.  The slightly longer merger time of PSR 1913+16 than
expected from the density of systems in this parameter space is
explained in section 4.1 of the text.}
\label{fig:tv}
\end{figure} 
Figure \ref{fig:tv} shows the distribution of merger times as a
function of system velocity. A majority of systems merge in $\sim
10^8$ yr spread over system velocities of $50$ --- $500$ km s$^{-1}$.
A subclass of systems are have spatial velocity and merger time which
are anti-correlated.

\begin{table*}
\label{table:sd}
\centering
\caption{The spatial distribution of coalescing neutron stars in
various galactic potentials. Though the average distance from center a
pair travels before coalescence ($d_{\rm avg}$) generally decreases
with increasing galactic mass, the median distance ($d_{\rm median}$)
scales with disk radius ($r_{\rm disk}$)}.
\vspace{0.2cm} 
  \begin{tabular}{lccccccl}
\hline\hline
 & \multicolumn{5}{c}{Galaxy parameters} & \multicolumn{2}{c}{Coalescence Distance} \\
Run & $v_{\rm circ}$ (km/s) & $r_{\rm break}$ (kpc) & $r_{\rm
  disk}$ (kpc) &  $M$ ($10^{11} M_{\odot}$) & L & $d_{\rm median}$ (kpc) & $d_{\rm avg}$ (kpc) \\
\hline\hline 
 a &  100 &  1  & 1 & 0.092 & $\ale  0.05 L_{*}$ & 4.3 & 66.2 \cr
 b &  100 &  3  & 1 & 0.278 & $\simeq  0.1 L_{*}$ & 4.0 & 50.1 \cr 
 c &  100 &  3  & 3 & 0.278 & $\simeq  0.1 L_{*}$ & 8.7 & 68.8 \cr
 d &  150 &  3  & 1 & 0.625 & $\simeq  0.5 L_{*}$ & 3.1 & 24.8 \cr
 e &  150 &  3  & 3 & 0.625 & $\simeq  0.5 L_{*}$ & 7.7 & 54.1 \cr
 f &  225 &  3  & 3 &  1.41 & $\simeq  1 L_{*}$  & 6.0  & 29.9 \cr
 g &  225 &  3  & 1 &  1.41 & $\simeq  1 L_{*}$  &  2.3 & 7.1 \cr
 h &  225 &  5  & 3 &  2.34 & $\simeq  2 L_{*}$  &  6.0 & 21.4 \cr
 i &  225 &  5  & 5 &  2.34 & $\simeq  2 L_{*}$  &  9.9 & 30.2 \cr
\hline
  \end{tabular} 
\end{table*}

\subsection{Coalescence/Birth Rates}

We have explored the consequences of different kick strengths on the
birthrates of NS--NS binaries. Table \ref{table:rate} summarizes these
results.

Earlier work (eg.~Sutantyo 1978; Dewey \& Cordes 1987; Verbunt, Wijers
\& Burm 1990; Wijers et al.~1992; Brandt \& Podsiadlowski 1995) in
which asymmetric kicks were incorporated with a single NS component
binary (as in LMXBs, HMXBs) noted a decrease in birthrate with
increased kick magnitude.  Portegies Zwart \& Spreeuw (1996) and
Lipunov, Postnov, \& Prokhorov (1997) found a similar effect on the
bound NS pair birthrates.  Lipunov (1997) provides a good review of
the expected rates. Clearly the birthrate of NS--NS binaries is also
sensitive to the total SN type II rate (which is observationally
constrained to no better than a factor of two, and theoretically
depends both on the uncertain high mass end of the initial mass
function and the total star formation at high redshift), and is also
sensitive to the fraction of high mass stars in binaries with high
mass secondaries.

We concentrate our discussion of NS--NS binary birthrates to galactic
systems for which the SN type II is fairly well-known (such as in the
Galaxy). It is important to note, however, that the SN type II rate
may be quite high in low surface-brightness and dwarf galaxies
(eg.~Babul \& Ferguson 1996). This would subsequently lead to a higher
NS--NS birthrates in such systems than a simple mass scaling to rates
derived for the Galaxy.

Recently van den Heuvel \& Lorimer (1996) find (observationally) the
birthrate of NS--NS binaries to be 8 Myr$^{-1}$. Lipunov, Postnov, \&
Prokhorov (1997) find between 100 and 330 events per Myr in
simulations.  Portegies Zwart \& Spreeuw (1996) found birthrates
anywhere from 9 to 384 Myr$^{-1}$ depending mostly on the choice of
asymmetric kick strength in their models.  We note that our derived
birthrate of $\sim 3$ Myr$^{-1}$ for high $\sigma_v = 270$ km s$^{-1}$
is comparable to those found Portegies Zwart \& Spreeuw (particularly
model ``ck'') with an average 3-D kick velocity of $450$ km s$^{-1}$.
Also, for low velocity kicks ($\sigma_{\rm kick} = 95$ km s$^{-1}$)
our birthrates approach that of Portegies Zwart \& Spreeuw models with
no asymmetric kicks.

The discrepancies between this and other work, therefore, we believe
are largely due to the choices of supernovae kick distributions and
strengths. That the absolute birthrate varies by an order of magnitude
depending on the binary evolution code and asymmetric kick
distributions used in different studies, hints at the uncertainty in
the knowledge of the true birthrates.
\begin{table*}
\label{table:rate}
\centering
\caption{The bound NS--NS binary birthrate and merger time properties
as a function of supernova kick strength.  A Maxwellian distribution
characterized by a velocity dispersion ($\sigma_{\rm kick}$) is
assumed. }
\vspace{0.2cm} 
  \begin{tabular}{cccl}
\hline\hline
$\sigma_{\rm kick}$ (km/s) & Birthrate (Myr $^{-1}$) & 
	$\tau_{\rm median}$ (yr) & $\tau_{\rm avg}^a$ (yr) \\
\hline \hline
 95  &  49 & $1.4 \times 10^8$  & $9.4 \times 10^8$ \cr
 190 &  10 & $7.0 \times 10^7$ &  $8.0 \times 10^8$ \cr
 270 &  3  & $5.5 \times 10^7$  & $7.0 \times 10^8$ \cr
\hline
\end{tabular} 

\raggedright $^a$ Average merger time of pairs merging in less than
  $1.5\times 10^{10}$ years.
\end{table*}

\subsection{Spatial Distribution}

\begin{figure}
\centerline{\psfig{file=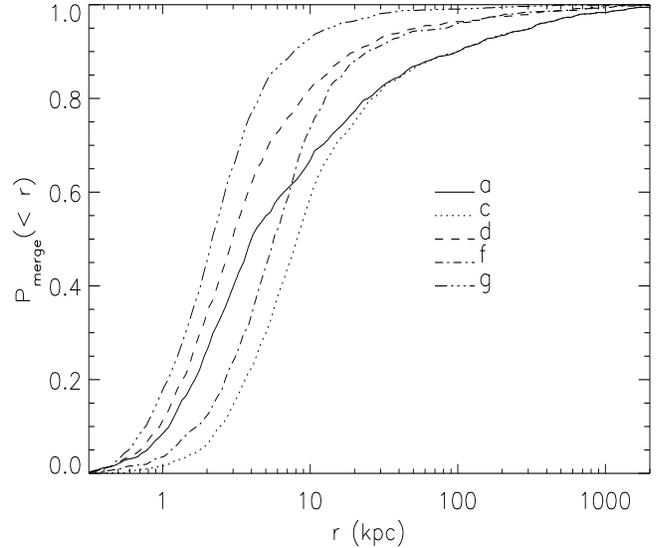,height=3.2in,width=3.8in}}
\caption[]{The radial distribution of coalescing neutron stars around
galaxies of various potentials.  The letters refer to runs in table 1.
In all scenarios, at least 50\% of the mergers occur within 10 kpc of
the host galaxy.  The wider radial distribution of in the
underluminous galaxy scenarios (a,c) reflects the smaller
gravitational potential of underluminous galaxies. }
\label{fig:face}
\end{figure}

Approximately half of the NS--NS binaries merge within $\sim 10^8$
years after the second SN; this merger time is relatively quick on
the timescale of star-formation. In addition, half the pairs coalesce
within a few kpc of their birthplace and within 10 kpc of the galactic
centre (see figure \ref{fig:face}) {\it regardless} of the potential
strength of the host galaxy.  As shown in figure \ref{fig:face},
galaxies with $M_{\rm galaxy} > 10^{10} M_\odot$ ($L \age 0.1 L_*$),
90 (95) percent of the NS--NS mergers will occur within 30 (50) kpc of
the host.
In the least massive dwarf galaxies with $M_{\rm galaxy} \simeq 9
\times 10^{9} M\odot$ ($\ale 0.1 L_*$), 50 (90, 95) percent of mergers
occur within $\sim 10$ ($100$, $300$) kpc of the host (see figure
\ref{fig:face}).  So, for example, assuming a Hubble constant of $H_0
= 65$ km s$^{-1}$ Mpc$^{-1}$ and $\Omega = 0.2$, we find that 90 (95)
percent of NS binaries born in dwarf galaxies at redshift $z=1$ will
merge within $\sim 12.7$ arcsec ($\sim 38.2$ arcsec) of the host
galaxy.  These angular offsets can be considered the extreme of the
expected radial distribution since the potentials are weakest and we
have not included the effect of projection.  We would expect 50 (90,
95) percent the mergers near non-dwarf galaxies to occur within $\sim
1.3$ (3.8, 6.4) arcsec from their host at $z=1$ for the cosmology
assumed above.

Given the agreement of our orbital parameter distribution (figure 1)
and velocity distribution (figure 2) with that of Portegies Zwart \&
Yungelson (1998), the discrepancy between the derived spatial
distribution (see figure 8 of Portegies Zwart \& Yungelson) is likely
due to our use of a galactic potential in the model.  This inclusion
of a potential naturally keeps merging NSs more concentrated towards
the galactic centre than without the effect.

\section{Discussion}


Although the NS--NS birthrate decreases with increased velocity kick,
the distribution of merger time and system velocity is not affected
strongly by our choice of kick distributions. Rather, the shapes of
the orbital and velocity distributions (figures 1 and 2) are closely
connected with the pre-SN orbital velocity, which is itself connected
simply with the evolution and masses. That is, bound NS binaries come
from a range of parameters which give high orbital velocities in the
pre-second SN system.  The orbital parameters (and merger time
distribution) of binaries which survive the second SN are not
sensitive to the exact kick-velocity distribution.  We suspect this
may be because bound systems can only originate from a parameter space
where the kick magnitude and orientation are tuned for the pre-second
SN orbital parameters.  The overall fraction of systems that remain
bound {\it is} sensitive to the kick distribution insofar as the kick
distribution determines how many kicks are in the appropriate range of
parameter space.
\begin{figure}
\centerline{\psfig{file=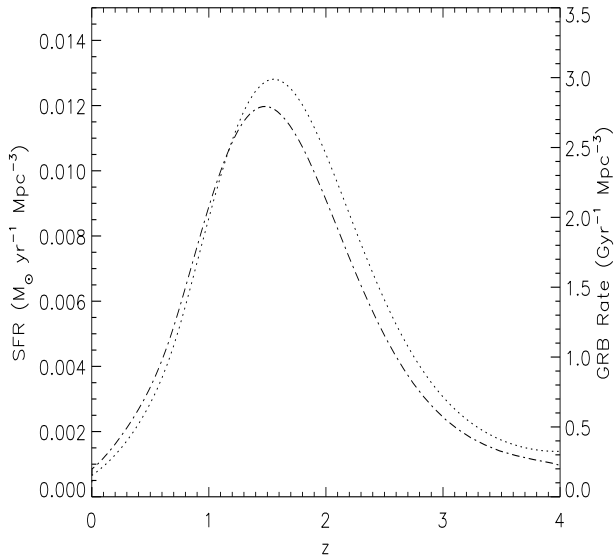,height=3.2in,width=3.5in}}
\caption[]{NS--NS merger rate dependence on redshift. The dotted line
is a reproduction of the SFR from Madau 1997 with
corresponding units on the left--hand axis.  The SFR curve as seen as a
lower limit to the true star-formation history since dust may obscure
a large fraction SFR regions in galaxies. The right hand axis is the
(unobscured) GRB rate if the bursts arise from the merger of two
NS--NS assuming a merger time distribution found in the present study
(dot--dashed line).  Both the SFR and merger rate are in co-moving
units (assuming $H_0 = 50$ km s$^{-1}$ Mpc$^{-1}$).  The normalisation
of the burst rate is taken from Wijers et al.~1998.}
\label{fig:sfr}
\end{figure} 

Since NS--NS binaries are formed rapidly (with an average time since
ZAMS of $\sim 22$ million years) and the median merger time is of
order one hundred million years regardless of the kick velocity
distribution (see table 2), the rate of NS--NS mergers should closely
trace the star formation rate.  In the context of gamma-ray bursts,
where merging NSs are seen as the canonical production mechanism, this
result implies that the GRB merger rate should evolve proportionally
to the star formation rate (see figure \ref{fig:sfr}; see also Bagot,
Portegies Zwart, \& Yungelson 1998).  Indeed, several studies (Totani
1997; Lipunov, Postnov, \& Prokhorov 1997; Wijers et al.~1998) have
consistently fit the GRB log $N$--log $P$ curve to a model which
assumes such a rate density evolution.

If indeed gamma-ray bursts arise from the coalescence of neutron star
binaries, then we confirm that GRBs should trace the star formation
rate in the Universe; thus most GRBs should have redshifts near the
peak in star formation (currently believed to be $1 \ale z \ale 2$;
Madau et al.~1996) although the observed distribution may be skewed to
lower redshifts by obscuration at high redshift (eg.~Hughes et
al.~1998). Determination of the distribution of x--ray and optical
counterparts to GRBs may help constrain the true cosmological star
formation history, though the observations of GRB counterparts are
vulnerable to some of the same extinction selection effects that
complicate determination of high redshift star formation rates.
Figure \ref{fig:sfr} illustrates the redshift dependence of the GRB
rate assuming the bursts arise in NS mergers.

The minimum required local (isotropic) bursting rate of 0.025 galactic
event per Myr (Wijers et al.~1998) is consistent with our birthrate
results (table 2) assuming a beaming fraction of 1/10--1/100 for the
gamma ray emission and our canonical values for the type II supernova
rate and supernova binary fraction.  The effects of beaming should be
observed in both the light curves of GRB afterglow and in deep
transient searches (eg.~Woods \& Loeb 1998).

In the case of GRB 970508, Bloom et al.~1998a and Castro-Tirado et
al.~1998 (see also Natarajan et al.~1998) found that the host is an
underluminous dwarf galaxy; the close spatial connection (offset $<
1$'') of the OT with the galaxy is then a case (albeit weakly) against
the NS--NS merger hypothesis as the {\it a priori} probability is
$\ale 20\%$ (figure \ref{fig:face}).  Paczy\'nski 1998 first pointed
out that the close spatial association with a dwarf galaxy is a case
against the NS--NS merger hypothesis.  Certainly more transients are
required to rule against the NS--NS merger hypothesis; we note,
however, that dust obscuration and projection effects may severely
bias the sample (see above discussion).

The verdict on the reconciliation of the expected radial distribution
of NS--NS mergers with hosts of known GRBs is still out.  Sahu et
al.~1997 found the optical transient associated with GRB 970228 to be
slightly offset from the centre of a dim galaxy but without a redshift
it is still unclear as to the the true luminosity of the host and thus
the expected offset of the OT in the NS--NS merger hypothesis.
Similarly, small or negligible offsets of GRB afterglow with faint
galaxies has been found for GRB 980326 and GRB 980329 (Djorgovski et
al.~1998).  Kulkarni et al.~1998 found the redshift of the purported
host galaxy of GRB 980329 to be $z=3.4$ implying the host is $L \age
L_*$; the expected offsets of NS--NS mergers around massive galaxies
(figure \ref{fig:face}, models d through i) is then consistent with
their finding of an OT offset $\simeq$ 0.5 kpc.

A few well-established offsets cannot tell us what is the true
distribution of GRBs around host galaxies.  As more OTs are
discovered, we will hopefully build up a large sample to statistically
test statistically the offsets.  Fortunately the unobscured afterglow
emission strength is coupled with the density $n$ of the surrounding
interstellar medium (ISM) with intensity scaling as $\sqrt n$
(Begelman, M\'esz\'aros \& Rees 1993; M\'esz\'aros, Rees, \& Wijers
1998); however, high ISM densities tracing dust will tend to obscure
rest-frame UV and optical emission from the transient.  In the absence
of strong absorption from the surrounding medium transients of GRBs
are preferential found close to where they are born, in the disk.
However, dust obscuration and projection effects severely complicate
determination of the true offset of OTs from their host galaxy.
Furthermore, identification of the host with a GRB becomes
increasingly difficult with distances beyond a few light radii ($\sim
10$ kpc) of galaxies (although see Bloom, Tanvir \& Wijers 1998).

If all afterglows, especially those where little to no absorption is
implied, are found more highly concentrated than predicted in figure
\ref{fig:face}, the NS--NS merger hypothesis would lose favour to
models which keep progenitors more central to their host. GRBs as
events associated with single massive stars such as microquasars
(Paczy\'nski 1998) or failed type Ib SN (Woosley 1993) could be
possible.  Alternatively, one may consider neutron star--black hole
(BH) binaries as the progenitors of GRBs (Mochkovitch et al.~1993;
M\'esz\'aros \& Rees 1997).  Most black hole X-ray binaries have
low-spatial velocities (although Nova Sco has $v_{\rm sys} \simeq 100$
km s$^{-1}$; see Brandt, Podsiadlowski, \& Sigurdsson 1995) so NS--BH
binaries should have system velocities $\sim 3$ to 10 times smaller
than NS--NS binaries.  One would expect NS--BH systems to be borne
with higher eccentricities than NS--NS systems leading to quicker
merger. Moreover, NS--BH binaries are more massive than NS--NS
binaries and merger time due to gravitational radiation scales
strongly with mass. Thus the attraction is that NS--BH mergers would
be preferentially closer to their host and their merger rate might be
small enough so as to require no beaming. Alternatively, gamma-ray
bursts could arise from several of these plausible progenitor models
and still be consistent with basic relativistic fireball models.

\section{Conclusions}

A reconciliation with the expected distribution of presumed
progenitors of GRBs and observed transient/host offsets is clearly
required.  We find for all plausible galactic potentials that the
median radial offset of a NS--NS merger is less than 10 kpc.  And in
all but the most shallow potentials, ninety percent of NS--NS binaries
merge within 30 kpc of the host. At a redshift of $z=1$ (with $H_0 =
65$ km s$^{-1}$ Mpc$^{-1}$ and $\Omega = 0.2$), this means that ninety
percent the coalescences should occur within $\sim 4$ arcsec from the
host galaxy.  Although the expected spatial distribution of coalescing
neutron star binaries found herein is consistent with the close
spatial association of known optical afterglows of gamma-ray bursts
with faint galaxies, a non-negligible fraction ($\sim 15$\%) of GRBs
{\it should} occur well outside dwarf galaxy hosts if the NS--NS
hypothesis is correct.  Otherwise, other models which keep progenitors
closer to their host (eg.~BH--NS mergers, microquasars, or ``failed SN
type Ib'') would be preferred.

As all the progenitor models mentioned are connected with high-mass
stars, the true GRB afterglow rate as a function of redshift should
trace the star-formation rate in the Universe.  However, environmental
effects, such as dust obscuration, may severely bias the estimate of
the true offset distribution.  Even in the NS--NS models where
progenitors have a natural mechanism to achieve high spatial
velocities, most will be closely connected spatially to their
host. Redshifts derived from absorption in the afterglow spectra
should be nearly always that of the nearest galaxy (Bloom et
al.~1998b).  Rapid burst follow-up ($\ale$ 1 hr), with spectra taken
while the optical transients are bright should confirm some form of
absorption from the host galaxy.

We have confirmed the strong dependence of birthrate of NS--NS
binaries on kick velocity distribution and found the independence of
the orbital parameters after the second supernova (and hence merger
times and spatial velocity) on the choice of kicks.  The methodology
herein can be extended to include formation scenarios of black
holes. This could provide improved merger rate estimates for LIGO
sources, and estimate the relative contribution of coalescences
between neutron stars and low mass black holes to the event
rate. Detailed modeling of the Milky Way potential would also allow
predictions for the distribution of NS--PSR binaries observable in the
Milky Way, which would provide an independent test of the assumptions
made in these models.

\vspace{-.5cm}
\section*{Acknowledgments}
It is a pleasure to thank Peter M\'esz\'aros, Melvyn Davies, Gerald
Brown, Hans Bethe, Ralph Wijers, Peter Eggleton, Sterl Phinney, Peter
Goldreich, Brad Schaefer, and Martin Rees for helpful insight at
various stages of this work. We especially thank Simon Portegies Zwart
as referee. JSB thanks the Hershel Smith Harvard Fellowship for
funding. SS acknowledges the support of the European Union through
a Marie Curie Individual Fellowship.


\begin{thebibliography}{99}

\bibitem[]{}Babul, A., Ferguson, H.~C.~1996, ApJ, 458, 100.

\bibitem[]{}Bagot, P., Portegies Zwart, S.~F., Yungelson, L.~R.~1998,
A\&A, in press.

\bibitem[]{}Begelman, M.~C., M\'esz\'aros, P., and Rees, M.~J.~1993,
MNRAS 265, L13.

\bibitem[]{}Blaauw, A.~1961, BAN, 15, 265.

\bibitem[]{}Bloom, J.~S., Djorgovski, S.~G., Kulkarni, S.~R., Frail,
D.~A.~1998a, ApJ Lett., submitted (astro-ph/9807315).

\bibitem[]{}Bloom, J.~S., Sigurdsson, S., Wijers, R.~A.~M.~J, Almaini,
O., Tanvir, N.~R., Johnson, R.~A.~~1998b, MNRAS, 292, 55L.

\bibitem[]{}Bloom, J.~S., Tanvir, N.~R., Wijers, R.~A.~M.~J.~1998,
(astro-ph$/$9705098).

\bibitem[]{}Brandt, N., Podsiadlowski, P.~1995, MNRAS, 274, 461.

\bibitem[]{}Cordes, J.~M., Chernoff, D.~1997, ApJ, 482, 971.

\bibitem[]{}Costa, E.~et al.~1997 IAU Circular~No.~6572.

\bibitem[]{}Dewey, R.~J., Cordes, J.~M.~1987, ApJ, 321, 780

\bibitem[]{}Djorgovski, S.~G.~et al.~1998, ApJL, submitted.

\bibitem[]{}Eichler, D., Livio, M., Prian, T., Schramm, D.~N.,
1989, Nature, 340, 126.

\bibitem[]{}Fenimore, E.~E., Bloom, J.~S.~1995, ApJ, 453, 25.

\bibitem[]{}Fryer, C., Burrows, A., Benz, W.~1997, ApJ, accepted
(Steward Observatory preprint 1416).

\bibitem[]{}Goodman, J.~1986, ApJ, 308, L17.

\bibitem[]{}Hartman, J.~W.~1997, A\& A, 322, 127.

\bibitem[]{}Hansen, B.~M.~S, Phinney, E.~S.~1997, MNRAS, 291, 569.

\bibitem[]{}Hernquist, L.~1990, ApJ, 356, 359 .

\bibitem[]{}Hughes, D.~et al.~1998, Nature, 394, 241.

\bibitem[]{}Johnston, H.~M., Kulkarni, S.~R.~1991, ApJ, 368, 504.

\bibitem[]{}Katz, J.~I.~1994, ApJ, 432, L107.

\bibitem[]{}Kulkarni, S.~R.~et al.~1998, Nature, 393, 35.

\bibitem[]{}Kulkarni, S.~R., Bloom, J.~S., Frail, D.~A., Ekers, R.,
Wieringa, M., Wark, R., Higdon, J.~L.~1998, IAU Circular, 6903.

\bibitem[]{}Lipunov, V.~M., Postnov, K.~A., Prokhorov, M.~E.,
Panchenko, I.~E., Jorgensen, H.~E, 1995, ApJ, 454, 593.

\bibitem[]{}Lipunov, V.~M., Postnov, K.~A., Prokhorov, M.~E.~1997,
MNRAS, 288,245.

\bibitem[]{}Lipunov, V.~M.~1997, astro-ph/9711270.

\bibitem[]{}Lyne, A.~G., Lorimer, D.~R.~1994, Nature, 369, 127

\bibitem[]{}Madau, P.~et al.~1997, MNRAS, 283, 1388

\bibitem[]{}M\'esz\'aros, P., Rees, M.~1997, ApJ, 476, 232.

\bibitem[]{}Metzger, M.~R., Djorgovski, S.~G., Steidel, C.~C.,
Kulkarni, S.~R., Adelberger, K.~L., Frail, D.~A.~1997, IAU
Circular 6655.

\bibitem[]{}M\'esz\'aros, P., Rees, M.~J.,~1993, ApJ, 418, 59.

\bibitem[]{}M\'esz\'aros, P., Rees, M.~J.,~1997, ApJ, 482, 29.

\bibitem[]{}M\'esz\'aros, P., Rees, M.~J., Wijers,
R.~A.~M.~J.~1998, ApJ, submitted (astro-ph$/$9709273).

\bibitem[]{}Mochkovitch, R., Hernanz, M., Isern, J., Martin, X., 1993,
Nature, 361, 236.

\bibitem[]{}Narayan, R., Ostriker, J.~P.~1990, ApJ, 352, 222.

\bibitem[]{}Narayan, R., Paczy\'nski, B., Piran, T.~1992, ApJ, 395,
83.

\bibitem[]{}Narayan, R., Piran, T., Shemi, A.~1991, ApJ, 397, L17.

\bibitem[]{}Natarajan, P.~et al.~1997, New Astronomy, 2, 471.

\bibitem[]{}Paczy\'nski, B.~1986, ApJ, 308, L43.

\bibitem[]{}Paczy\'nski, B.~1990, ApJ, 358, 485.

\bibitem[]{}Paczy\'nski, B.~1998, ApJ Letters, 494, 45.

\bibitem[]{}Paczy\'nski, B.~ \& Rhoads, J.~1993, ApJ, 418, L5.

\bibitem[]{}Peters, P.~C.~1964, Phys. Rev., 136, 1224

\bibitem[]{}Phinney, E.~1991, ApJL, 380, L17.

\bibitem[]{}Piro, L., Scarsi, L., Butler, R.~C.~1995, Proc.~SPIE
2517, 169-181.

\bibitem[]{}Pols, O.~R., Marinus, M.~1994, A\& A, 288, 475.

\bibitem[]{}Pols, O.~R.~1994, A\&A, 290, 119.

\bibitem[]{}Portegies Zwart, S.~F., Spreeuw, H.~N.~1996, A\&A, 312, 670

\bibitem[]{}Portegies Zwart, S.~F., Verbunt, F.~1996, A\&A, 309,
179.

\bibitem[]{}Portegies Zwart, S.~F., Yungelson, L.~1998, A\& A,
332. 173.

\bibitem[]{}Rees, M.~J., M\'esz\'aros, P. 1992, MNRAS, 258, 41.


\bibitem[]{}Rees, M.~J., M\'esz\'aros, P. 1994, ApJ, 430, 93.

\bibitem[]{}Sari, R., Prian, T., Narayan, R.~1998, ApJ, 497, 17.

\bibitem[]{}Sahu, K.~C.~et al.~1997, ApJ Lett., 429, 127.

\bibitem[]{}Sutantyo, W.~1978, Ap \& SS, 54, 479

\bibitem[]{}Tammann, G.~A., Loffler, W., Schrofer, A., 1994, ApJS,
92, 487.

\bibitem[]{}Taylor, J.~R., Dewey, R.~J.~1988, ApJ, 332, 770.

\bibitem[]{}Taylor, J.~R., Weisberg, J.~M.~1989, ApJ, 345, 434.

\bibitem[]{}Totani, 1997, ApJ, 486, L71.

\bibitem[]{}Tutukov, A.~V., Yungelson, L.~R.~1994, MNRAS, 268,
871.
 
\bibitem[]{}van den Heuvel, E.~P.~J., Lorimer, D.~R.~1996, MNRAS,
283, L37.

\bibitem[]{}Verbunt, F., Wijers, R.~A.~M.~J., Burm, H.~M.~G.~1990,
A\&A, 234, 195

\bibitem[]{}Van Paradijs, J.~et al.~1997, Nature, 386, 686. 

\bibitem[]{}Waxman, E.~1997, ApJ, 489, 33.

\bibitem[]{}Waxman, E., Kulkarni, S.~R., Frail, D.~A.~1998, ApJ,
497, 288.

\bibitem[]{}Wettig, T., Brown, G.~E.~1996, New Astronomy, 1, 17-34.

\bibitem[]{}Vietri, M. 1997, ApJ, 478, L9.

\bibitem[]{}Wijers, R.~A.~M.~J., van Paradijs, J., van den Heuvel,
E.~P.~J.~1992, A \& A, 261, 145.

\bibitem[]{}Wijers, R.~A.~M.~J., Rees, M.~J., M\'esz\'aros, P., MNRAS,
288, L51.

\bibitem[]{}Wijers, R.~A.~M.~J, Bloom, J.~S., Natarajan, P., 
Bagla, J.~S.~1998, MNRAS, 294, 13.

\bibitem[]{}Wolszczan, A.~1991, Nature, 350, 688.

\bibitem[]{}Woods, E., Loeb, A.~1998, ApJL, submitted.

\bibitem[]{}Woosley, S.~1993, ApJ, 405, 273.

\bibitem[]{}Yungelson, L., Portegies Zwart, S.~F., 1998, astro-ph$/$9801127.

\end{thebibliography}
\end{document}